\documentclass[showpacs,preprintnumbers,amsmath,amssymb,floatfix]{revtex4}

\usepackage{color}   
\usepackage{graphicx}
\usepackage{dcolumn}
\usepackage{bm}

\begin{document}
\def\bbox#1{\hbox{\boldmath${#1}$}}
\def\blambda{{\hbox{\boldmath $\lambda$}}}
\def\eeta{{\hbox{\boldmath $\eta$}}}
\def\bxi{{\hbox{\boldmath $\xi$}}}
\def\bzeta{{\hbox{\boldmath $\zeta$}}}

\title{ Ridge Structure associated with the Near-Side Jet in the
$\Delta\phi$-$\Delta \eta$ Correlation }

\author{Cheuk-Yin Wong}

\affiliation{Physics Division, Oak Ridge National Laboratory, 
Oak Ridge, TN\footnote{wongc@ornl.gov} 37831}

\date{\today}

\begin{abstract}

In the $\Delta \phi$-$\Delta \eta$ correlation associated with a
near-side jet observed by the STAR Collaboration in heavy-ion
collisions at RHIC [Ref. 1-6], the ridge structure can be explained by
the momentum kick model in which the ridge particles are identified as
medium partons which suffer a collision with the jet and acquire a
momentum kick along the jet direction.  If this is indeed the correct
mechanism, the ridge structure associated with the near-side jet may
be used to probe the parton momentum distribution at the moment of the
jet-parton collision, leading to the result that at that instant the
parton temperature is slightly higher and the rapidity width
substantially greater than corresponding quantities of their evolution
product inclusive particles at the end point of the
nucleus-nucleus collision.
\end{abstract}

                                                                         
\maketitle

\section {Introduction}

Recently, the STAR Collaboration
\cite{Ada05,Ada06,Put07,Bie07,Bie07a,Lon07} observed a $\Delta
\phi$-$\Delta \eta$ correlation of particles associated with a jet in
RHIC collisions, where $\Delta \phi$ and $\Delta \eta$ are
respectively the azimuthal angle and pseudorapidity differences
measured relative to a trigger jet particle.  The experimental
measurements, as shown in Fig. 1$a$ and discussed in detail in Refs.\
\cite{Ada05,Ada06,Put07,Bie07,Bie07a,Lon07}, reveal the following
features of the ``ridge phenomenon'':

\begin{enumerate}
\item 
There are particles associated with the trigger jet within a small
cone of $(\Delta \phi,\Delta \eta)\sim (0, 0)$ which belong to
the remnants of the near-side jet component. The near-side jet yield
is independent of the number of participants.
\item 
In addition, there are particles associated with the trigger jet
within a small range of $\Delta \phi$ around $\Delta \phi = 0$ but
distributed broadly in $\Delta \eta$.  These associated particles can
be separated as the ridge component associated with the trigger jet.
\item
The yield of the ridge component increases approximately linearly with
the number of participants and is nearly independent of (i) the flavor
content, (ii) the meson/hyperon character, and (iii) the transverse
momentum $p_t$ (above 4 GeV) of the jet trigger
\cite{Put07,Bie07,Bie07a}.  The ridge yield has a temperature that is
similar (but slightly higher) than that of the inclusive yield,
whereas the near-side jet component has an appreciably higher
temperature.
\end{enumerate}

\begin{figure} [h]
\includegraphics[angle=0,scale=0.50]{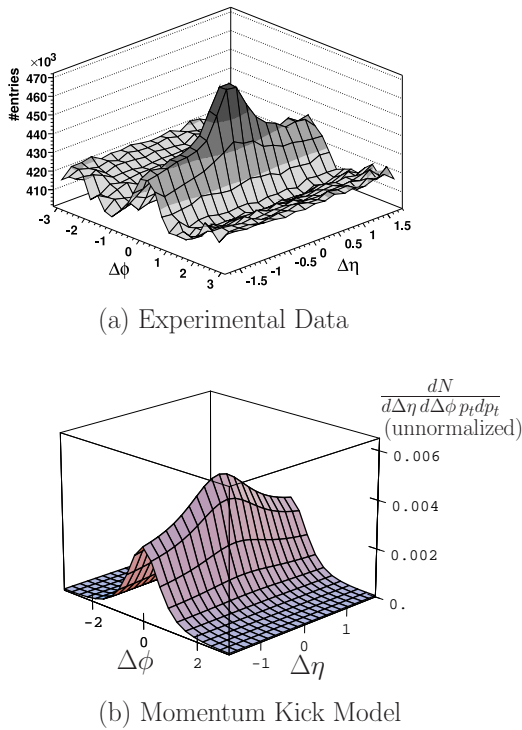}
\vspace*{-3.0cm}
\caption{ (a) The experimental yield of associated particles as a
function $\Delta \phi$ and $\Delta \eta$ \cite{Put07}, showing both
the near-side jet at $(\Delta \phi,\Delta \eta) \sim (0,0)$ and the
ridge structure at $\Delta \phi\sim 0$ and $|\Delta \eta| > 0.5$.  (b)
The unnormalized yield $dN/d\Delta \eta d\Delta \phi p_t dp_t$ in the
momentum kick model for the description of only the ridge component,
calculated for $p_t=2$ GeV with $q=0.8$ GeV, $\sigma_y=5.5$, and
$T=0.47$ GeV.  }
\end{figure}

While many theoretical models have been proposed to discuss the jet
structure and related phenomena
\cite{Hwa03,Chi05,Rom07,Vol05,Arm04,Shu07,Pan07}, the ridge phenomenon
has not yet been fully understood.  We would like to propose a simple
``momentum kick model'' which gives the salient features of the ridge
phenomenon.  We would also like to discuss the related implications of
using the ridge structure to probe the parton momentum distribution at
the moment of the jet-parton collision.

This paper is organized as follows.  In Section II, we introduce the
momentum kick model which involves the momentum kick along the jet
direction and the parton momentum distribution at the moment of the
jet-parton collision.  In Section III, we discuss the parametrization
of the initial parton momentum distribution.  In Section IV, we
examine the effects of the momentum kick and the initial rapidity
width on the final momentum distribution after the jet-parton
collision.  Adopting the hadron-parton duality, we compare the
experimental data with momentum kick model results in Section V.  In
Section VI, we study the ridge structure with identified particles. In
Section VII, we examine other predictions of the momentum kick model.
In Section VIII, we present the conclusions and discussions.

\section{The Momentum Kick Model}

Experimental observation (3) above concerning the properties of the
ridge particles implies that these particles are associated with
partons in the produced dense medium and much less with (i) the flavor
content, (ii) the meson/hyperon character, and (iii) the transverse
momentum $p_t$ (above 4 GeV) of the trigger jet.  We are led to such a
suggestion of medium partons because the associated ridge particles
bear the same ``fingerprints'' as those of the inclusive particles,
including their various particle yields increasing with the
participant number and their having nearly similar temperatures.
These characteristics suggest that the ridge particles originate from
the same material as the produced partons of the dense medium, which
later cool down and materialize also as inclusive particles.

Experimental observation (2) above regarding the narrow $\Delta \phi$
correlation of the ridge particles with the jet implies that these
ridge particles of the produced dense medium acquire their azimuthally
properties from the jet.  The most likely explanation is that these
ridge particles are medium partons which suffer a collision with the
jet and acquire a momentum kick and the jet's directionality.

Therefore, to describe the ridge distribution, we propose a simple
momentum kick model that incorporates the essential elements of the
experimental observations.  The momentum kick model contains the
following ingredients:

\begin{enumerate}

\item
A near-side jet occurs near the medium surface and the jet collides
with partons in the medium on its way to the detector.

\item
The jet-parton collision samples the momentum distribution of the
collided medium partons at the moment of the jet-parton collision.
Because of the condition for the occurrence of the near-side jet,
these collided medium partons are near the surface and each collided
parton suffers at most one collision with the jet.

\item
The jet-parton collision imparts a momentum kick ${\bf q}$ to the
collided parton in the direction of the jet.  The momentum kick
modifies the parton initial momentum distribution $P_i(\bbox{p}_i)$ to
turn it into the collided parton final momentum distribution
$P_f(\bbox{p}_f)$.  After the collided partons hadronize and escape
from the surface without further collisions, they materialize as ridge
particles which retain the collided parton final momentum
distribution.

\end{enumerate}

In mathematical terms, the final momentum distribution of the medium
partons which suffer a collision with the jet is,
\begin{eqnarray}
\label{eq1}
P_f({\bf p}_f)= \int \frac{d{\bf p}_i}{E_i} \int d {\bf q} ~P_i({\bf
p}_i) P_q ({\bf q}) ~E_f \delta ({\bf p}_f-{\bf p}_i-{\bf q}).
\end{eqnarray}
By normalizing the momentum kick distribution $P_q({\bf q})$ as
\begin{eqnarray}
\int d{\bf q}P_q({\bf q})=1,
\end{eqnarray}
the kinematic quantities $E_i$ and $E_f$  in Eq.\ (\ref{eq1}) ensure the
conservation of collided parton numbers before and after the
jet-parton collision,
\begin{eqnarray}
N_f=  \int \frac{d{\bf p}_f}{E_f} ~P_f({\bf p}_f)
= \int \frac{d{\bf p}_i}{E_i} ~P_i({\bf p}_i)=N_i.
\end{eqnarray}
In the context of the present paper, the initial and final moment
distributions $P_i(\bbox{p}_i)$ and $P_f(\bbox{p}_f)$ refer to the
momentum distributions before and after the jet-parton collision
respectively, not to be confused with the initial and final momentum
distributions at other instants of the nucleus-nucleus collision.

The momentum kick distribution $P_q(\bbox{q})$ for the momentum
$\bbox{q}$ imparted from the jet to a collided medium parton is not a
quantity that can be obtained from rigorous first-principles of QCD at
present, as many non-perturbative properties of the medium and of the
collision process are not known.  For the present analysis, of
particular interest are the general features of this momentum kick
that a collided parton experiences.  We know that this momentum kick
will be distributed within a narrow cone along the jet direction.
While a future refinement to describe this cone distribution is
possible, it is illuminating at this stage to introduce as few
parameters as possible for the description of this momentum kick, in
order to spell out the dominant effects of the jet-parton collision
process.  Accordingly, we shall use a single parameter $q$ to describe
the momentum kick distribution simply as
\begin{eqnarray}
P_q ({\bf q}) = \delta (\bbox{q}-q \bbox{e}_{jet}),
\end{eqnarray}
where $q=|{\bf q}|\ge 0$, and $\bbox{e}_{jet}$ is the unit vector
along the jet direction.  The simplicity of the model allows us to
obtain analytically from Eq.\ (\ref{eq1}) the distribution of partons
after suffering a jet-parton collision as
\begin{eqnarray}
\label{pi}
P_f({\bf p}_f) &=&~\left [ P_i({\bf p}_i)
\frac{E_f}{E_i} 
\right ]_{\bbox{p}_i=\bbox{p}_f-q \bbox{e}_{jet}}
\nonumber\\ &=&~\left [ P_i({\bf p}_i)
\frac{\sqrt{m^2+p_{f}^2}}
{\sqrt{m^2+p_{i}^2}}
\right ]_{\bbox{p}_i=\bbox{p}_f-q \bbox{e}_{jet}}.
\end{eqnarray}

\section{Parametrization of the parton momentum distribution}

In Eq.\ (\ref{pi}), the initial and final parton momentum
distributions can be presented in terms of Cartesian momentum
components in the collider frame, $\bbox{p}=(p_{1},p_{2},p_{3})$, with
a longitudinal $p_3$ component, a transverse $p_1$ component, and a 
$p_2$ component perpendicular to both the $p_1$ axis and the $p_3$
axis.  The coordinate axes can be so chosen that the trigger jet lies
in the $p_1$-$p_3$ plane.  The parton momentum distributions can
alternatively be represented in terms of the rapidity $y$, the
transverse momentum $\bbox{p}_t$, and the azimuthal angle $\phi$.
They are related to each other.  By changing the variables from the
Cartesian components $\bbox{p}_f=(p_{f1},p_{f2},p_{f3})$ to
$(y_f,\phi_f,p_{tf})$, we have
\begin{eqnarray}
 P_f(p_{f1},p_{f2},p_{f3}) 
=\frac{ E_f~dN_f~ }{ dp_{f1}dp_{f2}dp_{f3}}
=\frac{dN_f}{dy_f d\phi_f p_{tf}dp_{tf}}.
\end{eqnarray}
Similarly, for the initial parton momentum distribution we have
\begin{eqnarray}
 P_i(p_{i1},p_{i2},p_{i3}) 
=\frac{dN_i}{dy_i d\phi_i p_{ti}dp_{ti}}.
\end{eqnarray}
The momentum kick model with a simple delta function momentum kick
distribution then gives from Eq.\ (\ref{pi})
\begin{eqnarray}
\label{pf}
\frac{dN_f}{dy_f d\phi_f p_{tf}dp_{tf}} 
=\left [ \frac{dN_i}{dy_i d\phi_i p_{ti}dp_{ti}} 
\frac{E_f}{E_i} \right ]_{\bbox{p}_i=\bbox{p}_f-q \bbox{e}_{jet}}.
\end{eqnarray}
The experimental data are presented not as a rapidity distribution but
as a pseudorapidity distribution of the final particles.  We need to
convert $dN_f/dy_f...$ to $dN_f/d\eta_f...$ \cite{Won94} and we have
\begin{eqnarray}
\label{eq9}
\frac{dN_f}{d\eta_f d\phi_f p_{tf}dp_{tf}} 
=\frac{dN_f}{dy_f d\phi_f p_{tf}dp_{tf}} 
\sqrt{1-\frac{m^2}{m_{tf}^2 \cosh^2 y_f}},
\end{eqnarray}
where $m$ is the rest mass of the parton and
$m_{tf}=\sqrt{m^2+p_{tf}^2}$.

The above analytical results allow us to evaluate the momentum
distribution of the collided partons.  We also expect that as the
near-side jet occurs near the medium surface, the collided partons
also reside near the surface.  The energetic collided partons will
likely not suffer additional collisions before they hadronize and
emerge from the surface as ridge particles.  The ridge particles
should possess the properties of the energetic collided parton final
momentum distribution.  We shall therefore assume hadron-parton
duality and identify the momentum $\bbox{p}$ of the associated
particles as the final momentum $\bbox{p}_f$ of the energetic partons
after the jet-parton collision in Eqs.\ (\ref{pf}) and (\ref{eq9}).
This will allow us to identify the distribution $dN_f/d\eta_f d\phi_f
p_{tf}d p_{tf}$ of Eq.\ (\ref{eq9}) as the observed distribution
$dN/d\eta d\phi p_{t}d p_{t}$ of associated ridge particles, to be
compared with experimental data.

The parton-hadron duality is a reasonable description for the
hadronization of energetic partons.  We can consider an energetic
quark parton [or a $(\bar q \bar q)_3$ color-triplet cluster parton of
two antiquarks] of various flavors and masses.  The quark parton [or
the $(\bar q \bar q)_3$ cluster parton] can pick up an antiquark from
the sea, which arises spontaneously in a quark-antiquark fluctuation
in the medium, and can subsequently emerge out of the medium to become
a colorless hadron.  Similarly, an energetic antiquark parton [or a
$(qq)_{\bar 3}$ cluster parton of two quarks] can pick up a sea quark
and can subsequently emerge as a colorless hadron. On the other hand,
an energetic gluon parton can turn itself into a color-octet $(q\bar
q)_8$ pair which can subsequently emit a soft gluon to become a
colorless color-singlet $(q\bar q)_1$ quarkonium when it emerges from
the medium.  For energetic partons, the loss of momentum and energy in
picking up a sea quark or antiquark, or emitting a soft gluon, can be
approximately neglected. The result is then the (approximate)
parton-hadron duality for energetic partons we are considering in the
present analysis.

We shall describe the initial momentum distribution of partons on the
right-hand side of Eq.\ (\ref{pf}) by a Gaussian distribution in
rapidity $y_i$ with a width parameter (the standard deviation)
$\sigma_y$, a thermal transverse momentum distribution characterized
by a `temperature' $T$, and a uniform azimuthal distribution in
$\phi_i$,
\begin{subequations}
\begin{eqnarray}
\label{dis1}
 \frac{dN_i}{dy_i d\phi_i p_{ti}dp_{ti}}&=&A_i
e^{-y_i^2/2\sigma_y^2} 
\frac{ \exp \{ -\sqrt{m^2+p_{ti}^2}/T \}} {\sqrt{m^2+p_{ti}^2}},
\\
\label{dis2}
A_i &=&\frac{N_i e^{m/T}} {(2\pi)^{3/2}\sigma_y T} ,
\end{eqnarray}
\end{subequations}
where $N_i$ is a total number of partons that collide with the jet.
Note that the `initial parton temperature' $T$ has been introduced
using the above form of the transverse momentum distribution.  This
functional form has been chosen to give a simple description of the
experimental central collision inclusive transverse momentum spectrum
from 0.2 GeV to $\sim$4 GeV, characterized by an inclusive
`temperature' $T_{incl}=0.40$ GeV [see Fig. 5(a) below].  To be
consistent, we expect the initial parton temperature $T$ introduced
here to be higher than the temperature $T_{incl}$ of inclusive
particles, as the initial partons will presumably cool down and evolve
to become the inclusive particles at the end-point of parton
evolution.

For a trigger jet with a transverse momentum $p_{t jet}$ at a
pseudorapidity $\eta_{jet}$, we have
\begin{eqnarray}
|{\bf p}_{jet}|=p_{t jet} \cosh \eta_{jet},
\end{eqnarray}
and
\begin{eqnarray}
p_{3 jet}=p_{t jet} \sinh \eta_{jet}.
\end{eqnarray}
As a consequence, the unit vector along the jet direction is
\begin{eqnarray}
{\bf e}_{jet}=\frac{{\bf e}_1 + \sinh \eta_{jet} {\bf e_3}}
{\cosh \eta_{jet}},
\end{eqnarray}
where ${\bf e}_1$ and ${\bf e}_3$ are unit vectors along the $p_1$ and
$p_3$ directions, respectively.  The final momentum distribution is
therefore
\begin{eqnarray}
\label{final}
\frac{dN_f}{d\eta_f d\phi_f p_{tf}dp_{tf}} 
=\left [ A_i e^{-y_i^2/2\sigma_y^2} \frac{ e^{-m_{ti}/T}}{m_{ti}}
\frac{E_f}{E_i} \right ]_{\bbox{p}_i=\bbox{p}_f-q \bbox{e}_{jet}}
\sqrt{1-\frac{m^2}{m_{tf}^2 \cosh^2 y_f}},
\end{eqnarray}
where the initial momentum $\bbox{p}_i=(p_{i1},p_{i2},p_{i3})$ is
related to the final momentum $\bbox{p}_f=(p_{f1},p_{f2},p_{f3})$ and
the trigger jet rapidity $\eta_{jet}$ by
\begin{subequations}
\begin{eqnarray}
p_{i1}&=&p_{f1}-\frac{q}{\cosh \eta_{jet}},\\
p_{i2}&=&p_{f2},\\
p_{i3}&=&p_{f3}-\frac{q\sinh \eta_{jet}}{\cosh \eta_{jet}},
\end{eqnarray}
\end{subequations}
and $y_i$ and $m_{ti}$ are the initial rapidity and transverse mass
\begin{eqnarray}
y_i=\frac{1}{2} \ln~\frac{E_i+p_{i3}}{E_i-p_{i3}},
\end{eqnarray}
\begin{eqnarray}
m_{ti}=\sqrt{m^2 + p_{i1}^2 + p_{i2}^2}.
\end{eqnarray}

The number of partons that collide with the jet $N_i$ is an overall
normalization constant. At this stage, we shall examined
``unnormalized'' distributions $dN/d{\Delta \eta}d{\Delta
\phi}p_tdp_t$ with $N_i$ set to 1 in Figs. 1, 2, 3, and 8, so that the
results can be rescaled when normalized experimental data are
available.  For numerical purposes, we take $m=m_\pi$, the mass of the
pion, unless indicated otherwise.

\section{Dependence of the Ridge Structure on the momentum Kick and the 
Rapidity Width}

Experimental data are presented in terms of $\Delta \eta=\eta -
\eta_{jet}$ and $\Delta \phi = \phi-\phi_{jet}$, relative to the
trigger jet $\eta_{jet}$ and $\phi_{jet}$.  We can obtain $dN/d\Delta
\eta d\Delta \phi p_t dp_t$ relative to the jet pseudorapidity and
azimuthal angle from $dN/d\eta d\phi p_t dp_t$ by a simple change of
variables.  For such a purpose, we need the experimental
pseudorapidity distribution of the trigger jet in STAR measurements,
\cite{Wan07}
\begin{eqnarray}
\label{dndeta}
\frac{dN_{jet}}{d\eta_{jet}}
=\frac{ \Theta (0.85- |\eta_{jet}|)}{1.7}.
\end{eqnarray}

We shall see in the next section that a reasonable set of parameters
that can reproduce the main features of the experimental data are
$q=0.8$ GeV, $\sigma_y=5.5$, and $T=0.47$ GeV.  To examine the effects
of the the momentum kick and the initial parton momentum distribution
on the $\Delta\phi$-$\Delta \eta$ correlation, we shall vary one of
the three parameters of $\{q,\sigma_y,T\}$ in this set, with the
remaining two parameters being held fixed.

\vspace*{+0.0cm}
\begin{figure} [h]
\includegraphics[angle=0,scale=0.50]{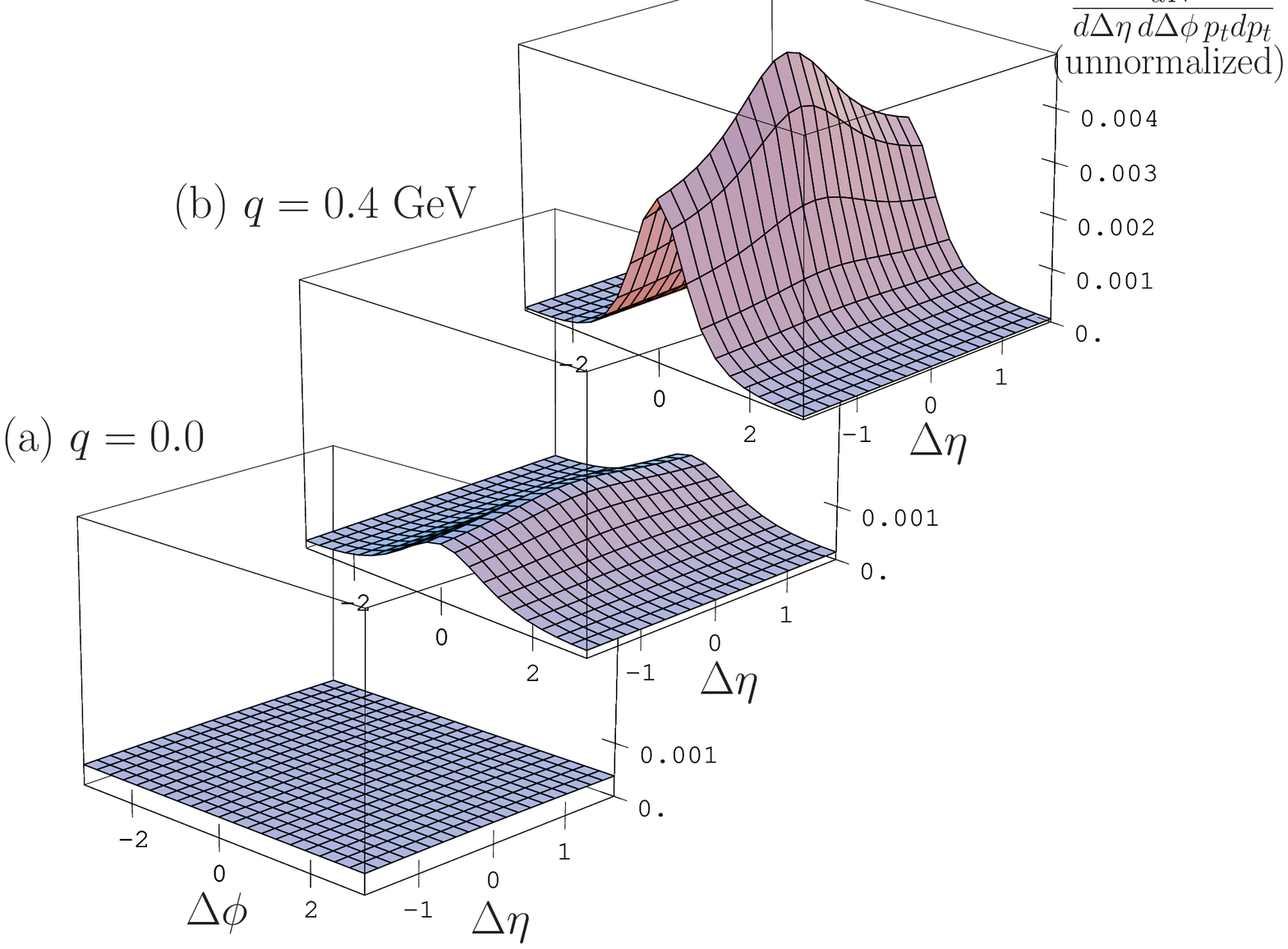}
\vspace*{-6.1cm}
\caption{ Th results of the the momentum kick model for the
unnormalized yield $dN/d\Delta \eta d\Delta \phi p_t dp_t$ of
associated ridge particles at $p_t= 2$ GeV, as a function $\Delta
\phi$ and $\Delta \eta$.  Figs. 2(a), 2(b), and 2(c) are for $q=0$,
0.4, and 0.8 GeV respectively.}
\end{figure}

We show in Fig.\ 2 the unnormalized yield $dN/d\Delta \eta d\Delta
\phi p_t dp_t$ of ridge particles at $p_t=2$ GeV obtained in the
momentum kick model for different values of $q$.  Figs. 2(a), 2(b),
and 2(c) are for $q=0$, 0.4, and 0.8 GeV respectively, the other
parameters being fixed as $\sigma_y=5.5$ and $T=0.47$ GeV.  They are
calculated with $N_i=1$ in Eqs.\ (\ref{dis1}) and (\ref{dis2}) for
$p_t= 2$ GeV.  As one observes in Fig.\ 2(a), when the momentum kick
is zero the distribution is essentially flat in $\Delta \phi$ and
$\Delta \eta$.  [There is a small variation of the distribution along
the $\Delta \eta$ direction, but with $\sigma_y=5.5$, the variation is
too small to show up in Fig.\ 2(a).] When there is a non-zero momentum
kick, the distribution develops a peak at $\Delta \phi = 0$ and a
ridge along the $\Delta \eta$ direction.  The height of the peak
increases and its width decreases, as the magnitude of $q$ increases.

It is easy to understand how a peak structure at $\Delta \phi=0$ along
the $\Delta \phi$ direction arises in the momentum kick model.
Consider a parton with a final momentum ${\bf p}_f=( \eta_f,\Delta
\phi,p_{tf})$ at a fixed pseudorapidity $\eta_f$ and transverse
momentum $p_{tf}$ with various azimuthal angles $\Delta \phi$.  Under
the action of a momentum kick along the jet direction, the square of
the magnitude of the initial parton transverse momentum $p_{ti}^2$ is
related to the final transverse momentum $p_{tf}$ by
\begin{eqnarray}
p_{ti}^2 =p_{tf}^2-2 p_{tf} ~q~ \cos\Delta \phi /\cosh \eta_{jet}
+q^2/\cosh^2 \eta_{jet}
\end{eqnarray}
The magnitude of the initial transverse momentum $p_{ti}$ is a minimum
at $\Delta\phi=0$, and it increases monotonically as $\Delta \phi$
increase to $\pi$. Because the initial transverse momentum $p_{ti}$ is
distributed according to $\exp\{-\sqrt{m^2+p_{ti}^2}
/T\}/\sqrt{m^2+p_{ti}^2}$, the probability distribution decreases for
increasing $p_{ti}$.  Thus, there are more collided partons at $\Delta
\phi=0$ than at $\Delta \phi =\pi$, for the same observed final
transverse momentum $p_{tf}$.  The case of azimuthal angles between
$\Delta \phi=0$ and $\Delta \phi=\pi$ gives a momentum distribution in
between these two limits, resulting in a peak structure in $\Delta
\phi$ centered at $\Delta\phi=0$.  The height of the peak increases
and its width decreases, as $q$ increases.

\vspace*{+0.0cm}
\begin{figure} [h]
\includegraphics[angle=0,scale=0.50]{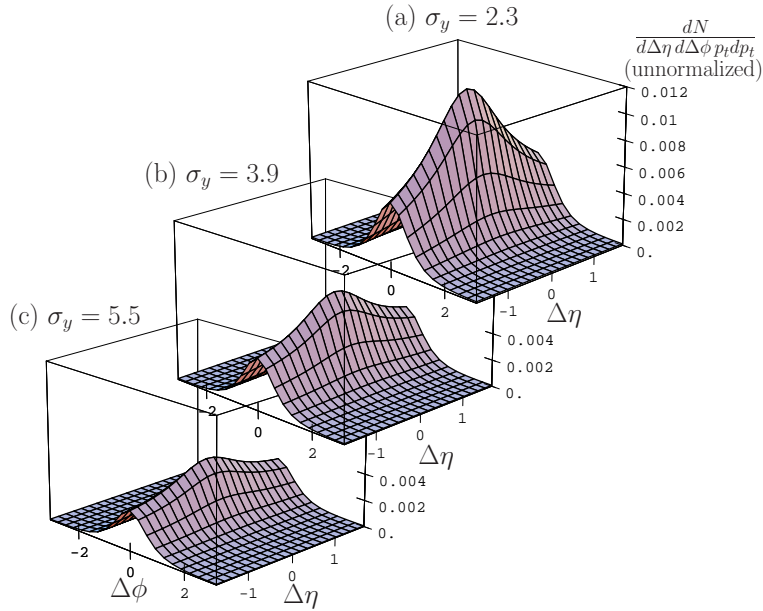}
\vspace*{-6.0cm}
\caption{ The results of the momentum kick model for the unnormalized
yield $dN/d\Delta \eta d\Delta \phi p_t dp_t$ of associated
particles at $p_t= 2$ GeV as a function $\Delta \phi$ and $\Delta
\eta$.  Figs. 3(a), 3(b), and 3(c) are for $\sigma_y=2.3$, 3.9, and
5.5 respectively.  }
\end{figure}

While the narrow peak as a function of $\Delta \phi$ depends mainly on
the momentum kick $q$, the ridge structure along the $\Delta\eta$
direction depends mainly on the initial rapidity distribution.  The
rapidity distribution of inclusive experimental data for central Au-Au
collisions at $\sqrt{s_{NN}}=200$ GeV can be well described by a
Gaussian distribution with a rapidity width parameter of
$\sigma_y=2.3$ which agrees with the Landau hydrodynamics prediction
of $\sigma_y=2.16$ within 5\% \cite{Mur04}. We would like to see
whether the ridge structure associated with the near-side jet can be
described by such a rapidity width parameter.  We would also like to
see how the ridge structure depends on the rapidity width parameter
$\sigma_y$.

Accordingly, we carry out calculations in the momentum kick model for
$\sigma_y=2.3$, 3.9, and 5.5 and the results of the unnormalized yield
$dN/d\Delta \eta d\Delta \phi p_t dp_t$ are shown in Figs. 3(a), 3(b) and 3(c)
respectively.  In these calculations, the other parameters are kept
fixed as $q=0.8$ GeV and $T=0.47$ GeV, and the results are obtained
for $p_t= 2$ GeV.

As one observes in Fig.\ 3(a), the distribution $dN/d\Delta \eta d\Delta \phi p_t
dp_t$ for $\sigma_y=2.3$ has a ridge with a steep slope toward the
maximum peak along the $\Delta \eta$ direction at $\Delta\phi=0$.  The
slope of the ridge appears to be too steep compared to the
experimental data (see also Fig.\ 4(b) below), when the distribution
is matched with the magnitude of the experimental data.  Apparently,
the width parameter that describes the inclusive rapidity distribution
data cannot reproduce the ridge structure associated with the jet.

The ridge becomes broader along the $\Delta \eta$ direction as
$\sigma_y$ increases.  When we increase the ridge width parameter to
$\sigma_y=3.9$ and $\sigma_y=5.5$, the slope of the ridge becomes less
steep as shown in Figs.\ 3(b) and 3(c).

\section{Comparison of the Model with Experimental Data}

We compare the momentum kick model results with the experimental data
\cite{Put07} at specific cuts in $\Delta \eta$ or $\Delta \phi$ in
Fig. 4 and with the two-dimensional $\Delta\phi$-$\Delta \eta$
distribution in Fig. 1. The experimental data contain both the
near-side jet and the ridge component, in addition to other unknown
contributions at $|\Delta \phi| > 1$.  The near-side jet component shows
up as a peak at $(\Delta \phi,\Delta \eta)\sim (0,0)$, and the
ridge component at $\Delta \phi \sim0$ with a relatively flat ridge
along the $\Delta \eta$ direction.

\begin{figure} [h]
\includegraphics[angle=0,scale=0.50]{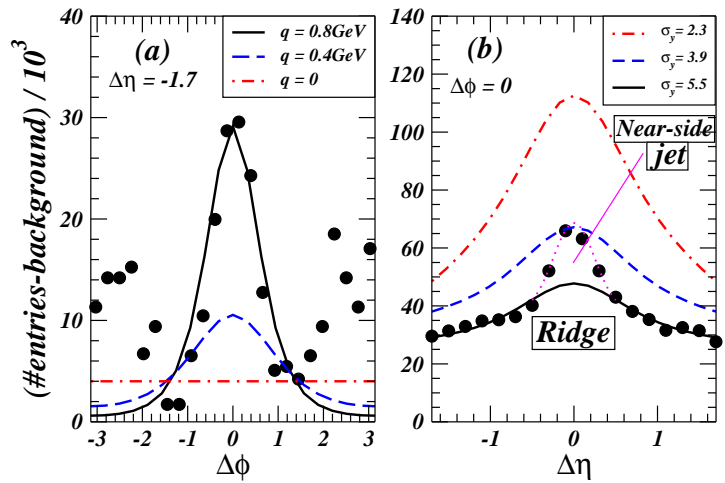}
\caption{ The unnormalized experimental associated particle data
\cite{Put07} shown as solid circular points, compared with the solid
curves calculated in the momentum kick model for $p_t=2$ GeV with
$q=0.8$ GeV, $\sigma_y=5.5$ and $T=0.47$ GeV.  The solid curves are
normalized by matching theoretical results with the experimental data
at $(\Delta \phi, \Delta \eta)=(0,-1.7)$.  Fig. 4(a) gives $dN/d\Delta
\eta d \Delta \phi p_t dp_t$ as a function of $\Delta \phi$ at $\Delta
\eta=-1.7$.    Fig.\ 4(b) gives $dN/d\Delta \eta d \Delta \phi p_t
dp_t$ as a function of $\Delta \eta$ at $\Delta \phi=0$.  The dashed,
and dash-dot curves are the momentum kick model results by varying $q$
in Fig. 4(a), and varying $\sigma_y$ in Fig. 4(b).  }
\end{figure}

As the ridge component has not been cleanly separated, we shall be
content at this stage with a qualitative analysis to compare the
results of the momentum kick model only with the ridge portion of the
experimental data \cite{Put07}, without considering the near-side peak
at $(\Delta \phi, \Delta \eta)\sim (0,0)$ and other contributions with
$|\Delta \phi| > 1$.

We find that the `best-fit' set of parameters $q = 0.8$ GeV,
$\sigma_y=5.5$, and $T=0.47 $ GeV gives a reasonable qualitative
description of the ridge component of the experimental data, as shown
by the solid curves in Figs.\ 4(a) and 4(b).  In these comparisons,
the theoretical calculations are normalized by matching the
experimental distribution \cite{Put07} at the point
$(\Delta\phi,\Delta\eta)=(0,-1.7)$ with the momentum kick model result
calculated with the set of parameters $q = 0.8$ GeV, $\sigma_y=5.5$,
and $T=0.47 $ GeV.  The experimental data points are taken from Ref.\
\cite{Put07} and shown in Figs.\ 1(a) and 4.  The experimental
background is taken to be $405\times10^3$ entries such that the lowest
data point, at $(\Delta \phi, \Delta\eta)=(1.2,-1.7)$, is nearly at
the background level.  The data in Fig. 4 have not been corrected for
$v_2$ elliptic flow.  The $v_2$ corrections involves the product of
$v_2($jet$)\sim 0.1$ and the $v_2($associated particles$)\sim 0.05$
and will probably modified the resultant parameters only slightly.  It
will be of interest to readjust the parameters when the
$v_2$-corrected data become available.

We study the variation of shape of the ridge structure as a function
of $q$ at $\Delta \eta =-1.7$ in Fig.\ 4(a), and separately as a
function of $\sigma_y$ at $\Delta \phi=0$ in Fig.\ 4(b). We calculate
the yield $dN/d\Delta \eta d\Delta \phi p_t dp_t$ at $p_t=2$ GeV for
values of $q$ and $\sigma_y$ different from those in the best-fit set,
using the same normalization as used for the best-fit set.  The
results of using parameters different from the best-fit set are shown
as the dashed and dash-dot curves in Figs.\ 4(a) and 4(b).  One notes
that many features of the theoretical results of these dashed and
dash-dot curves do not match those of the experimental data.  For
example, in Fig.\ 4(a) for $q=0.4$ GeV, the full width at half maximum
is about $\Delta\phi \sim 2.4$ which is much greater than the
experimental full width at half maximum of about $\Delta\phi \sim
1.2$.  In Fig.\ 4(b) for $\sigma_y=2.3$, the ridge yield increase by a
factor of about 1.7 when $\Delta \eta$ increases from -1.7 to -0.7,
whereas the experimental data shows an increase of only a factor of
about 1.2.  In contrast, the solid curves obtained with the best-fit
set of parameters, $q = 0.8$ GeV, $\sigma_y=5.5$, and $T=0.47 $ GeV,
give a good representation of the ridge data.

To compare the two-dimensional distributions, we plot in Fig. 1(a) the
experimental yield \cite{Put07} and in 1(b) the unnormalized yield
$dN/d\Delta \eta d\Delta \phi p_t dp_t$ in the momentum kick model, as
a function of both $\Delta \phi$ and $\Delta \eta$, at $p_t=$ 2 GeV
using our best-fit set of parameters, $q=0.8$ GeV, $\sigma_y=5.5$, and
$T=0.47$ GeV. The general two-dimensional features of the ridge are
qualitatively reproduced.  A more careful quantitative analysis of the
ridge structure will need to await quantitative data in which the
ridge component can be cleanly separated.  Ridge measurements
extending to greater values of $|\eta|$ will also be of great interest
in determining more accurately the rapidity width parameter $\sigma_y$
and finding out whether the initial rapidity distribution may be
non-Gaussian at the moment of the jet-parton collision.

\begin{figure} [h]
\includegraphics[angle=0,scale=0.50]{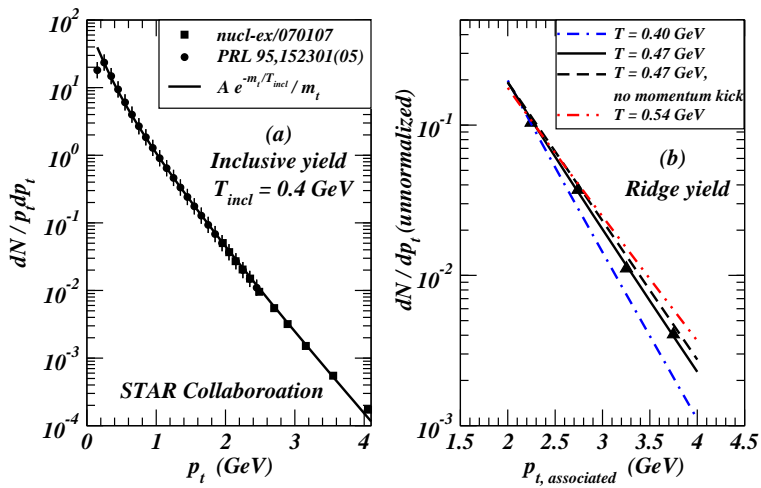}
\caption{ The central collision transverse momentum distributions of
(a) inclusive particles and (b) ridge particles.  The solid curve in
(a) is the theoretical distribution of $dN/p_t dp_t \propto
\exp\{-m_t/T_{incl}\}/m_t$ with $T_{incl}=0.40$ GeV, and the data
points are from Refs.\ \cite{Ada05} and \cite{Put07}.  In Fig. 5(b),
the data points are experimental ridge yield (proportional to
$dN/dp_t$) \cite{Put07}. The dash-dot, solid, and dash-dot-dot curves
are the $dN/dp_t$ results of the momentum kick model for $T=0.50,
0.47, 0.40$ GeV, respectively.  They are calculated with $q=0.8$ GeV
and $\sigma_y=5.5$ and normalized to match the data point at
$p_{t,assoc}=2.13$ GeV. The dashed curve is theoretical results of
$dN/dp_t \propto p_t\exp\{-m_t/T_{incl}\}/m_t$ with $T=0.47$ GeV and
no momentum kick.  }
\end{figure}

We can now examine the transverse momentum distribution of the ridge
particles.  We show in Fig. 5(a) and 5(b) the transverse distribution
data of inclusive particles \cite{Ada05} and ridge particles
\cite{Put07} respectively, together with theoretical fits.  The
inclusive $dN/p_t dp_t$ data in Fig. 5(a) can be described by a
distribution of the form $A\exp \{ -m_t/T_{incl} \}/m_t$ with an
inclusive `temperature' parameter $T_{incl}=0.40$ GeV.  The ridge
particle yield calculated in the momentum kick model for different
values of $T$ are given in Fig.\ 5(b) and compared with the
experimental transverse yield (proportional to $dN/dp_t)$ for a jet
trigger of $4 {\rm ~GeV} <p_{trig} <5 {\rm ~GeV}$ \cite{Put07}.  In
this comparison, we have normalized the results from the momentum kick
model by matching the theoretical $dN/dp_t$ results with the
experimental data point at $p_{t, associated}= 2.13$ GeV.  The
transverse momentum distribution from the momentum kick model with
$T=0.47$ GeV, represented by the solid curve, agrees well with data,
whereas the results from $T=0.40$ GeV and $T=0.54$ GeV given by the
dash-dot and dash-dot-dot curves do not agree with the ridge
transverse distribution.  As a further comparison, we show also the
theoretical transverse momentum distribution $dN/ dp_t \propto p_t
\exp\{-(m_t-m)/T\}/m_t$ with $T=0.47$ GeV without the momentum kick as
the dashed curve.  Comparison of the solid and the dashed curves
indicates that the shape of the transverse momentum distribution in
this $p_t$ region is changed only very slightly by the presence of the
momentum kick $q$.

The results of Fig.\ 5(a) and 5(b) indicate that the parton medium at
the moment of jet-parton collision has a temperature higher than the
temperature of the inclusive particles at the end point of parton
evolution, as noted earlier in a different parametrization of the
transverse momentum distribution \cite{Put07,Bie07,Bie07a}.

\section{Study of the Ridge Structure with Identified Particles}

Recent measurements with identified trigger particles \cite{Bie07} and
identified ridge particles \cite{Bie07a} provide useful information on
the ridge phenomenon.  The gross structure of the ridge is nearly
unaffected by the flavor, the meson/hyperon character, and the
transverse momentum (above 4 GeV) of the trigger jet.  This implies
that (i) the parton-jet interaction is approximately independent of
the jet structure, and (ii) the momentum kick $q$ suffered by the
parton in a jet-parton collision is, to the lowest order, a constant
quantity.  Finer dependencies of $q$ on the invariant mass and flavor
of the jet trigger are possible and may be explored by future accurate
measurements.

We shall first present some general results concerning the ratio of
local yields at a specific transverse momentum $p_t$ for identified
associated ridge particles within the momentum kick model.
Specifically, for two parton species with parton masses $m_1$ and
$m_2$ with parton numbers $N_1$ and $N_2$ , the momentum kick model
gives the final ratio at $p_t$ 
\begin{eqnarray} R_{m1/m2}(p_t)=\frac{dN_1/p_tdp_t (m_1,p_t) } {dN_2/p_tdp_t
(m_2,p_t) } \sim \frac{dN_{1}^{({\rm init})} /p_tdp_t (m_1,|p_t-q|) }
{dN_{2}^{({\rm init})}/p_tdp_t (m_2,|p_t-q|) }= R_{m1/m2}^{({\rm init})}
(|p_t-q|), \end{eqnarray} 
relating the observed ratio of the species at $p_t$ to the ratio of
the initial parton distributions at a lower momentum, $|p_t-q|$,
displaced by the momentum kick $q$.  If the $m_t$ scaling of the
initial parton distribution in the form of Eq.\ (10) is a good
description, then the initial transverse momentum distribution ratio
is given by
\begin{eqnarray}
R_{m1/m2}(p_t) &\sim&  \frac{dN_1^{({\rm init})}/p_tdp_t (m_1,|p_t-q|) } 
{dN_2^{({\rm init})}/p_tdp_t (m_2,|p_t-q|) } 
\nonumber \\
&=&
\frac{N_1}{N_2} 
\frac{e^{m_1/T}}{e^{m_2/T}}
\frac{\sqrt{m_2^2+(p_t-q)^2}}{\sqrt{m_1^2+(p_t-q)^2}}
\frac{\exp\{-\sqrt{m_1^2+(p_t-q)^2} /T\} }  
     {\exp\{-\sqrt{m_2^2+(p_t-q)^2} /T\} } .
\end{eqnarray}
For $p_t >> m_1, m_2$,
\begin{eqnarray} 
R_{m1/m2}(p_t)
\sim \frac{N_1}{N_2} e^{(m_1-m_2)/T},  
\end{eqnarray}
which is independent of $p_t$.  For $|p_t -q| \to 0$ (if the model
remains valid),
\begin{eqnarray} 
R_{m1/m2}(p_t) \sim \frac{N_1}{N_2} \frac{m_{2}}{m_{1}}.
\end{eqnarray}
It will be of interest to examine how these results can be confronted
with experimental measurements of different species among the ridge
particles.

\begin{figure} [h]
\includegraphics[angle=0,scale=0.50]{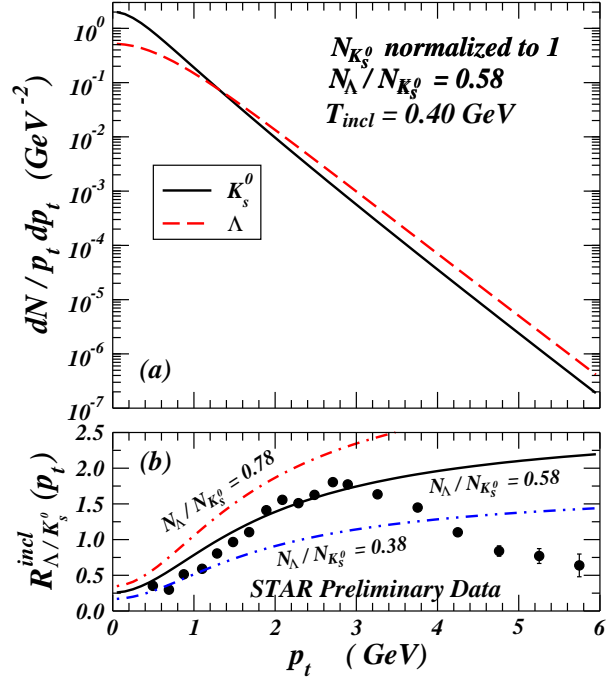}
\caption{ (a) Theoretical transverse momentum distributions of
inclusive $K_s^0$ and $\Lambda$ particles based on the distribution
Eq.\ (\ref{ptdis1}) with $N_{K_s^0}$ normalized to 1,
$N_\Lambda/N_{K_s^0}=0.58$, and $T_{inclusive}=0.40$ GeV.  Fig. 6(b)
gives the ratio of inclusive $R_{\Lambda/K_s^0}^{incl}(p_t)$ as a
function of $p_t$.  The dash-dot, solid, and dash-dot-dot curves are
the theoretical ratio $R_{\Lambda/K_s^0}^{incl}(p_t)$ calculated with
$N_{\Lambda}/N_{K_s^0}=0.78$, 0.58, and 0.38, respectively.  The
experimental data points are from Ref.\ \cite{Bie07a}.  }
\end{figure}

In the experiments performed by the STAR Collaboration, the ratio of
the yield of various identified particles have been obtained not at a
single $p_t$ but over a range of $p_t$ with trigger particles spanning
over a range of transverse momenta.  Specifically,  the experimental
$\Lambda/K_s^0$ ratio was obtained for Au on Au at $\sqrt{s_{NN}}=200$
GeV at 0-10\% centrality, within the kinematic domain specified by
\cite{Bie07a}
\begin{eqnarray}
\label{dom}
& & {p_{tjet,min}= 2{\rm ~ GeV}}, ~~~~ {p_{tjet,max}= 3{\rm ~ GeV}}, 
\nonumber \\
& & {\rm and~identified~ridge~particles~with~}p_{tmin}=1.5 {~\rm~ GeV~}<
p_{t}<p_{tjet}. 
\end{eqnarray}
In the momentum kick model, the ratio of the ridge particles
within the kinematic domain (\ref{dom}) as defined by the measurement
is given by
\begin{eqnarray} 
\label{lamk}
\langle R_{\Lambda / K_s^0} \rangle_{ridge} = \frac 
{\int_{p_{tjet,min}}^{p_{tjet,max}} P_{jet}(p_{tjet}) p_{tjet} dp_{tjet}
\int_{p_{tmin}}^{p_{tjet}} P_{\Lambda} (p_t - q) p_t dp_t}
{\int_{p_{tjet,min}}^{p_{tjet,max}} P_{jet}(p_{tjet}) p_{tjet} dp_{tjet}
\int_{p_{tmin}}^{p_{tjet}} P_{K_s^0} (p_t - q) p_t dp_t }.
\end{eqnarray}
We shall use hadron-parton duality and assume the equivalence of the
hadron mass and the parton mass.  The transverse momentum distribution
of the $\lambda$-type species in the above equation is given by
\begin{eqnarray} 
\label{ptdis}
P_\lambda(p_t)=\frac {dN_\lambda}{p_tdp_t}(p_t) = \frac{N_\lambda \exp
\{ -(\sqrt{p_t^2+m_\lambda^2}-m_\lambda)/T\}}
{\sqrt{p_t^2+m_\lambda^2}},
\end{eqnarray}
where $N_\lambda$ is the total number of the produced $\lambda$-type
parton in the medium,
\begin{eqnarray} 
N_\lambda=\int \frac {dN_\lambda}{p_tdp_t}(p_t)~~ p_tdp_t,
\end{eqnarray}
and $T$ is the initial parton temperature at the moment of jet-parton
collision, $T =0.47$ GeV.  For the jet trigger transverse momentum
distribution, we can use the experimental distribution
\begin{eqnarray} 
\frac{ dP} {p_{tjet} dp_{tjet}} \propto \exp\{-p_{tjet}/T_{jet} \}
\end{eqnarray}
with $T_{jet}=0.478$ GeV for hadron jet triggers \cite{Bie07a}.  [Note
that $T_{jet}$ pertains to the above form of $p_{tjet}$ distribution
\cite{Bie07a}, which differs from our form of the $p_t$ distribution,
Eq.\ (\ref{ptdis}), for initial partons.]

For the evaluation of the $\langle R_{\Lambda / K_s^0}
\rangle_{ridge}$ as given by Eqs.\ (\ref{lamk}) and (\ref{ptdis}), we
need to know the ratio of total parton numbers, $N_\Lambda/N_K$, in
the parton medium.  We can infer such a ratio from the inclusive data
where we have
\begin{eqnarray} 
R_{\Lambda / K_s^0}^{incl} (p_t)
=\frac{dN_\Lambda^{incl}/p_tdp_t (m_\Lambda,p_t) } 
{dN_{K_s^0}^{incl}/p_tdp_t (m_{K_S^0},p_t) },
\end{eqnarray}
and the inclusive yield is
\begin{eqnarray}
\label{ptdis1} 
\frac{dN_{\lambda}^{incl}}{p_tdp_t} (m_{\lambda},p_t) =
\frac{N_\lambda \exp\{-\sqrt{m_\lambda^2+p_t^2} /T_{incl}\} }
      {\sqrt{m_\lambda^2+p_t^2} }  .
\end{eqnarray}
The inclusive temperature parameter appropriate for the medium at the
end point of the parton evolution has been determined from Fig.\ 5(a)
to be $T_{incl}=0.40$ GeV.  In Fig. 6(a), we show the transverse
distributions of $K_s^0$ and $\Lambda$, normalized to $N_{K_s^0}=1$
and $N_\Lambda/N_{K_s^0} = 0.58$.  By taking the ratio of the two
types of particles at different transverse momenta in Fig.\ 6(a), we
obtain solid curve of $R_{ \Lambda / K_s^0}^{incl} (p_t)$ in Fig.\
6(b).  We also calculate $R_{ \Lambda / K_s^0}^{incl} (p_t)$ for the
ratios of $N_\Lambda/N_{K_s^0} = 0.38$ and 0.78, shown as the dash-dot
and dash-dot-dot curves in Fig.\ 6(b).  Comparison of $R_{ \Lambda /
K_s^0}^{incl} (p_t)$ experimental data points with the three different
predictions using different $N_\Lambda/N_{K_s^0}$ ratios indicates
that $N_\Lambda/N_{K_s^0} = 0.58$ gives a good description of the
experimental inclusive ratio $R_{ \Lambda / K_s^0}^{incl} (p_t)$ for
$p_t < 3 $ GeV.  The region of $p_t>3$ GeV involves additional
fragmentation mechanisms for baryon and meson production, and cannot
be described by the above equation of $m_t$ scaling.  Fortunately, the
kinematic domain of interest lies in the region of $p_t<3$ GeV.

Carrying out the integrations in Eq.\ (\ref{lamk}) and using the total
parton numbers ratio $N_\Lambda/N_{K_S^0}=0.58$ determined from
inclusive data of Fig.\ 6(b), we obtain $\langle R_{ \Lambda / K_s^0}
\rangle_{ridge} = 0.71$ within the kinematic domain specified by Eq.\
(\ref{dom}).  Within the experimental error, this theoretical ratio
for ridge particle in the momentum kick model is consistent with the
observed measurement of $\langle R_{\Lambda / K_s^0} \rangle_{ridge} =
0.81 \pm 0.14$.  The corresponding value for the jet is $0.46\pm
0.21$, which is smaller than $\langle R_{\lambda/K_s^0}
\rangle_{ridge} $.

\section{Other Predictions of the Momentum Kick Model}
 
The momentum kick model is based on simplifying idealizations.  The
model needs to be tested and improved by comparing its predictions
with experimental measurements.

We have assumed that the initial rapidity distribution is in the form
of a Gaussian distribution with a standard deviation $\sigma_y$.  A
distribution with $\sigma_y=5.5$ appears to give a good description of
the measured ridge pseudorapidity data up to $|\Delta \eta|<1.7$.  It is
also possible that the shape of the initial rapidity distribution may
be non-Gaussian or multi-center in nature.  It is therefore useful to
obtain the predicted distribution at large pseudorapidity, for a
Gaussian initial rapidity distribution.

For definiteness, we take the trigger jet distribution as given by
Eq.\ (\ref{dndeta}), corresponding to the experimental jet acceptance
of the STAR Collaboration around $\eta_{jet}\sim 0$.  We calculate the
pseudorapidity distribution of the associated particles for this
trigger jet distribution. The solid curves in Figure 7 shows the
predictions from the momentum kick model for $q=0.8$ GeV,
$\sigma_y=5.5$, and $T=0.47$ GeV, calculated for $p_t=2$ GeV.  The
fall-off of the ridge distribution is gradual and has a slope
different from that in the smaller $\Delta \eta$ region around
$\Delta\eta\sim 0$.  The decrease is rather slow as a function of
$\Delta \eta$, decreasing by about 30\% as $\Delta \eta$ changes from
$\Delta \eta =1.7$ to $\Delta \eta =5$.

\begin{figure} [h]
\includegraphics[angle=0,scale=0.50]{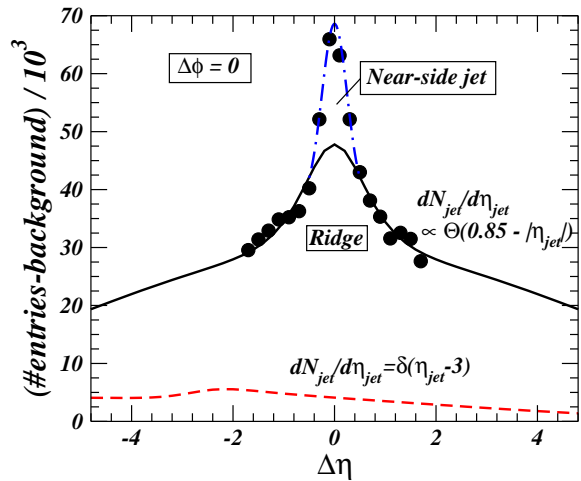}
\caption{ The yield $dN/d\Delta \eta d\Delta \phi p_t dp_t$ obtained
in the momentum kick model for the description of only the ridge
component, calculated for $p_t=2$ GeV with $q=0.8$ GeV,
$\sigma_y=5.5$, and $T=$ 0.47 GeV, normalized to the yield at $\Delta
\eta=-1.7$. The solid curve is for $dN_{\rm jet}/d\eta_{jet}=
\Theta(0.85-|\eta_{jet}|)/1.7$ and the dashed curve is for
$dN_{jet}/d\eta_{jet}=\delta(\eta_{jet}-3)$ }
\end{figure}

Previously, STAR has attempted to measure the associated particles at
forward rapidities.  They indeed observed a hint of the ridge effect
at forward rapidity although the systematic uncertainty is large
\cite{Mol07,Wanf07}.  It will be of great interest to make a
quantitative comparison so as to probe the initial rapidity
distribution in the forward region.

Another interesting question is to inquire what kind of associated
particle distribution will be expected for a forward jet trigger, say,
at $\eta_{jet}=3$.  The dashed curve in Fig. 6 shows the
pseudorapidity distribution of the associated particles as a function
of the pseudorapidity relative to the jet pseudorapidity for
$\eta_{jet}=3$.  We observe that the peak of the $dN/d\Delta \eta$
distribution of the associated particle is shifted and is located at
$\Delta \eta\sim - 2.2$ (corresponding to $\eta_{\rm Lab}\sim
0.8$). The shape of the distribution is not symmetrical with respect
to the peak and the magnitude of the distribution at the peak is
small compared with that for the case with the jet trigger at
$\eta_{jet}\sim 0$.

\begin{figure} [h]
\includegraphics[angle=0,scale=0.80]{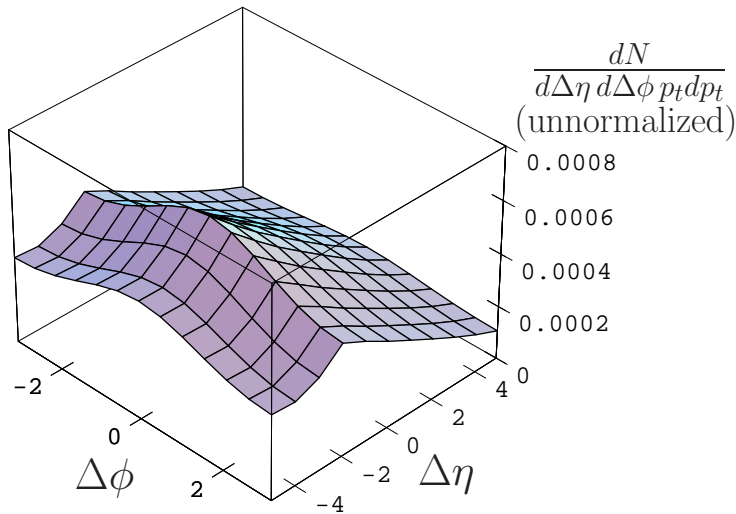}
\vspace*{-12.0cm}
\caption{ The unnormalized yield $dN/d\Delta \eta d\Delta \phi p_t
dp_t$ of the associated particles as a function of $\Delta\phi$ and
$\Delta \eta$ for a jet trigger of $\eta=3$, calculated with the
momentum kick model for $p_t=2$ GeV.  }
\end{figure}

\vspace*{0.0cm}
Figure 8 shows the 3-D associated particle distribution for a trigger
jet with $\eta_{jet}=3$, calculated for $p_t=2$ GeV with $q=0.8$ GeV,
$\sigma_y=5.5$, and $T=0.47$ GeV.  We observe that the ridge is now
not in the $\Delta \eta$ direction as in the case of jet trigger
$\eta\sim 0$ case, but in the $\Delta\phi$ direction.  The peak of the
ridge is shifted and located at $\Delta \eta\sim -2.2$, as mentioned
previously.  Although the collided parton gains a momentum kick along
the jet direction, the pseudorapidity of the peak is not at $\Delta
\eta=0$.  It is shifted and appears at $\Delta \eta\sim -2.2$ because
the longitudinal momentum imparted on the partons is limited and has a
magnitude of only $q=0.8$ GeV, in the direction of the jet. This
momentum kick shifts the parton longitudinal momentum of the partons
only by about 0.8 unit, resulting in a $\Delta \eta\sim -2.2$ relative
to the incident jet of $\eta_{jet}=3$.

Another question can be raised on the behavior of low-$p_t$ partons in
the momentum kick model.  We expect low-$p_t$ partons to be present at
the moment of jet-parton collision and these partons can collide with
the jet.  Under the action of a momentum kick, the initial parton
momentum is related to the final parton momentum by
\begin{eqnarray}
{\bf p}_{i}={\bf p}_{f}-q{\bf e}_{jet}.
\end{eqnarray} 
For simplicity, we can focus our attention at final partons at
$(\Delta \phi_f, \Delta \eta_f)\sim (0,0)$ with a trigger jet at
$\eta_{jet}\sim 0$, as other kinematic regions can be similarly
analyzed. Then, the magnitude of the initial parton transverse
momentum is
\begin{eqnarray}  { p}_{ti}=|{p}_{tf}-q|, \end{eqnarray} 
which has a minimum of zero at ${p}_{tf}=q$ and
it increases for $p_{tf}> q$ and $p_{tf}< q$.  In the momentum kick
model, the final momentum distribution is given by the momentum
distribution of the initial parton. One therefore expects that the
final (observed) transverse momentum distribution of ridge particles
has a peak at $p_{ft}= q$, decreasing on both sides for $p_{tf}>q$ and
$p_{tf}<q$.  Because $q$ has been found to have the value of 0.8 GeV,
we therefore expect that the transverse momentum spectrum of ridge
particles associated with the near-side jet will have a peak at
$p_{tf}\sim 0.8$ GeV at $(\Delta \phi_f, \Delta \eta_f)\sim (0,0)$.

The above prediction of the momentum kick model is based on the
idealization of parton-hadron duality for which the momentum and
energy loss due to the action of picking up a sea quark or antiquark,
or emitting a soft gluon, can be approximately neglected.  For low
$p_t$ partons, these momentum and energy loss in the hadronization
process may need to be corrected and accounted for, leading to
modifications that will likely smear and shift the predicted peak of
the transverse momentum distribution at $p_{ft}\sim q$.  How the
hadron-parton duality will be modified for these low $p_t$ partons
will need to be investigated both experimentally and theoretically.
In this connection, it is interesting to note that the experimental
transverse spectra, $dN/dp_t$, of particles associated with the
near-side jet in the low $p_t$ region below 1 GeV in central
collisions is relatively flat, in contrast distinction to the
exponentially-increasing $dN/dp_t$ spectrum of inclusive particles and
the $dN/dp_t$ spectrum of particles associated with the away-side jet,
as shown in Fig. 3 of \cite{Ada05}.  These data of particles
associated with a near-side jet in \cite{Ada05} contain both the jet
and the ridge components.  It will be of great interest to separate
out the near-side jet and the ridge components for low $p_t$
associated particles, to see whether a ridge exists and whether there
is any evidence of a momentum kick suffered by the initial partons in
producing these observed hadrons with low transverse momenta.

\section{Conclusions and Discussions}

The STAR Collaboration has observed a $\Delta \phi$-$\Delta \eta$
correlation of particles associated with a jet in RHIC collisions
\cite{Ada05,Ada06,Put07,Bie07,Bie07a,Lon07}.  The correlations can be
decomposed into a near-side component as remnants of the trigger jet
at $(\Delta \phi, \Delta \eta)\sim (0,0)$ and a ridge particle
component at $\Delta\phi \sim 0$ with a broad ridge structure in
$\Delta \eta$.  These two components have very different
characteristics.  The ridge particles bear the same ``fingerprints''
as those of inclusive particles, including their particle yields
increasing with increasing participant numbers and having nearly
similar temperatures.  These characteristics suggest that the ridge
particles originate from the same material of the produced partons of
the dense medium, which later also materialize as inclusive particles.

The $\Delta \phi\sim 0$ correlation of the ridge particles with the
trigger jet suggests further that the ridge particles and the jet are
related kinematically.  It is therefore reasonable to propose the
momentum kick model in which the ridge particles are identified as
medium partons which suffer a collision with the jet and acquire a
momentum kick along the jet direction.  After the collided partons
materialize as ridge particles and escape from the surface without
additional collisions, they retain the collided parton final momentum
distribution.  The yield of the ridge particles associated with the
jet therefore will depend on the momentum kick and the initial
momentum distribution of the collided partons before the jet-parton
collision.

To gain important insights into the ridge phenomenon, we idealize the
momentum kick model by introducing as few parameters as possible.  The
relevant quantities are then the magnitude of the momentum kick $q$
along the jet direction imparted by the jet to the collided parton,
and the initial parton momentum distribution represented by the
rapidity width parameter $\sigma_y$ and the transverse momentum
temperature $T$.

Under the action of a jet-parton momentum kick, a peak structure in
$\Delta \phi$ arises.  For the same final parton transverse momentum
$p_{tf}$ and various $\Delta \phi$ values, the jet-parton collision
samples a smaller magnitude of initial parton transverse momentum
$p_{ti}$ at $\Delta \phi=0$ but it samples a larger magnitude of
initial parton transverse momentum $p_{ti}$ at $\Delta \phi=\pi$.  As
the initial transverse momentum distribution decreases rapidly with
the initial transverse momentum $p_{ti}$ as
$\exp\{-\sqrt{m^2+p_{ti}^2}/T \}/m_{ti}$, there are more collided
partons at $\Delta \phi=0$ than at $\Delta \phi =\pi$, for the same
observed final transverse momentum $p_{tf}$. The case of azimuthal
angles between $\Delta \phi=0$ and $\Delta \phi=\pi$ gives a momentum
distribution in between these two limits, resulting in a peak
structure in $\Delta \phi$ centered at $\Delta\phi=0$.

This width of the distribution along the $\Delta \phi$ direction is
sensitive to the magnitude of the momentum kick $q$.  A small value of
$q$ will lead to a broad structure in $\Delta\phi$. A large value of
$q$ will lead to a narrow structure in $\Delta\phi$.  The experimental
data suggest a a value of the momentum kick $q$ of 0.8 GeV.  As there
is much that is not known about the non-perturbative properties of the
medium and the collision processes, the experimental extraction of the
jet-parton momentum kick provides useful information concerning the
jet-parton collision and jet energy loss.

While the peak structure along the $\Delta\phi$ direction at $\Delta
\phi =0$ arises predominantly from the momentum kick, the ridge
structure along the $\Delta\eta$ direction in the $\Delta\phi$-$\Delta
\eta$ plane on the other hand reflects the initial rapidity
distribution $dN/dy$ at the momentum of the jet-parton collision.
Thus, if the momentum kick model is indeed the correct mechanism, the
ridge structure associated with the near-side jet may be used to probe
the parton momentum distribution at the moment of the jet-parton
collision.

Previous measurements of inclusive particle rapidity distribution at
RHIC for $\sqrt{s_{NN}}=200$ GeV gives a width parameter
$\sigma_y=2.3$ which however leads to too steep a ridge slope in the
momentum kick model to be favored by the experimental ridge
distribution.  The rapidity width $\sigma_y\sim 5.5$ extracted from the
momentum kick model appears to be greater than the rapidity width
extracted from the inclusive particle rapidity distribution.

It is important to note that the rapidity width $\sigma_y$ extracted
from the momentum kick model measures the rapidity width
$\sigma_y({\rm JPC})$ at the moment of the jet-parton collision, where
JPC stands for ``jet-parton collision''.  The jet-parton collision
occurs early in the nucleus-nucleus collision.  The rapidity width
$\sigma_y$ measured by the inclusive particle distribution is the
asymptotic rapidity width $\sigma_y(t\to \infty)$ at the end point of the
nucleus-nucleus collision.

\begin{figure} [h]
\includegraphics[angle=0,scale=0.50]{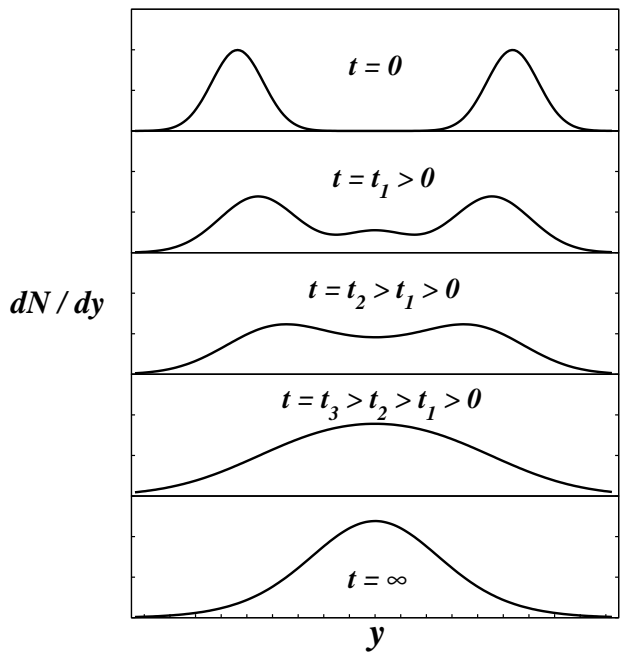}
\caption{ Possible evolution scenario of the parton rapidity
distribution $dN/dy$ as a function of time.
}
\end{figure}

Not much is known experimentally about the evolution of the rapidity
distribution of partons as a function of time.  Nevertheless, certain
gross features can perhaps be inferred from general considerations, as
depicted schematically in Fig. 9.  At the onset of the nucleus-nucleus
collision at $t=0$ in the collider frame, the partons have rapidity
distributions centered at the rapidities of the two colliding nuclei,
each distribution having a rapidity width because of the fermi motion
of the partons in the hadrons inside the nucleus.  At a subsequent
time $t=t_1$, the partons will thermalize and the rapidity
distribution will evolve from a two-center distribution into an
overlapping three-center distribution, with the thermalized component
at zero rapidity (Fig. 9).  The distributions on the two sides will
move towards zero rapidity and will diminish their magnitudes, while
the central distribution will gain in magnitude.  The three-center
distribution will eventually merge into a single distribution at times
$t=t_2$ and $t=t_3$ in Fig. 9.  In this scenario, the width
$\sigma_y({\rm JPC})$ should be larger than the asymptotic width
$\sigma_y(t\to \infty)$, as the rapidity distribution at the moment of the
jet-parton collision may only be in the early stage of the rapidity
evolution from the two-center distribution to an overlapping
three-center distribution and a merged single distribution.  Because
of the complicated dynamics that enters into the evolution of the
early rapidity distribution, it is possible that the rapidity
distribution at this stage may be non-Gaussian in shape.  The
determination of the full shape of the initial rapidity distribution
will need the measurement of associated ridge particles at large
values of $|\Delta\eta|$.

In conclusion, we find that the ridge structure in the $\Delta
\phi$-$\Delta \eta$ correlation associated with a near-side jet
\cite{Ada05,Ada06,Put07,Bie07,Bie07a,Lon07} can be explained in terms
of the momentum kick model in which the ridge particles are identified
as medium partons which suffer a collision with the jet and acquire a
momentum kick along the jet direction.  If this is indeed the correct
mechanism, the near-side jet may be used to probe the parton momentum
distribution at the moment of jet-parton collision, leading to the
result that at that instant the parton temperature is slightly higher
and the rapidity width substantially greater than corresponding
quantities of their evolution products of inclusive particles at the
end point of the nucleus-nucleus collision.

\vspace*{0.3cm} The author wishes to thank Drs. V. Cianciolo,
J. Putschke, Fuqiang Wang, and Gang Wang for valuable communications
and comments.  The author also wishes to thank Drs. Huan Huang,
C. Ogilvie, K. Read, and S. Sorensen for helpful discussions.  This
research was supported in part by the Division of Nuclear Physics,
U.S. Department of Energy, under Contract No.  DE-AC05-00OR22725,
managed by UT-Battle, LLC.

\end{document}